\documentclass[12pt]{article}
\usepackage{amssymb}

\def\hybrid{\topmargin 0pt      \oddsidemargin 0pt
        \headheight 0pt \headsep 0pt
        \voffset=-0.5cm
        \hoffset=-0.25in
        \textwidth 6.75in
        \textheight 9.5in       
        \marginparwidth 0.0in
        \parskip 5pt plus 1pt   \jot = 1.5ex}
\catcode`\@=11
\def\marginnote#1{}

\newcount\hour
\newcount\minute
\newtoks\amorpm
\hour=\time\divide\hour by60 \minute=\time{\multiply\hour by60 \global\advance\minute by-\hour}
\edef\standardtime{{\ifnum\hour<12 \global\amorpm={am}%
        \else\global\amorpm={pm}\advance\hour by-12 \fi
        \ifnum\hour=0 \hour=12 \fi
        \number\hour:\ifnum\minute<10 0\fi\number\minute\the\amorpm}}
\edef\militarytime{\number\hour:\ifnum\minute<10 0\fi\number\minute}

\def\draftlabel#1{{\@bsphack\if@filesw {\let\thepage\relax
   \xdef\@gtempa{\write\@auxout{\string
      \newlabel{#1}{{\@currentlabel}{\thepage}}}}}\@gtempa
   \if@nobreak \ifvmode\nobreak\fi\fi\fi\@esphack}
        \gdef\@eqnlabel{#1}}
\def\@eqnlabel{}
\def\@vacuum{}
\def\draftmarginnote#1{\marginpar{\raggedright\scriptsize\tt#1}}
\def\draftlabel#1{{\@bsphack\if@filesw {\let\thepage\relax
   \xdef\@gtempa{\write\@auxout{\string
      \newlabel{#1}{{\@currentlabel}{\thepage}}}}}\@gtempa
   \if@nobreak \ifvmode\nobreak\fi\fi\fi\@esphack}
        \gdef\@eqnlabel{#1}}
\def\@eqnlabel{}
\def\@vacuum{}
\def\draftmarginnote#1{\marginpar{\raggedright\scriptsize\tt#1}}

\def\draft{\oddsidemargin -.5truein
        \def\@oddfoot{\sl preliminary draft \hfil
        \rm\thepage\hfil\sl\today\quad\militarytime}
        \let\@evenfoot\@oddfoot \overfullrule 3pt
        \let\label=\draftlabel
        \let\marginnote=\draftmarginnote
   \def\@eqnnum{(\theequation)\rlap{\kern\marginparsep\tt\@eqnlabel}%
\global\let\@eqnlabel\@vacuum}  }

\def\numberbysection{\@addtoreset{equation}{section}
        \def\theequation{\thesection.\arabic{equation}}}

\def\underline#1{\relax\ifmmode\@@underline#1\else
        $\@@underline{\hbox{#1}}$\relax\fi}

\def\titlepage{\@restonecolfalse\if@twocolumn\@restonecoltrue\onecolumn
     \else \newpage \fi \thispagestyle{empty}\c@page\z@
        \def\thefootnote{\fnsymbol{footnote}} }

\def\endtitlepage{\if@restonecol\twocolumn \else  \fi
        \def\thefootnote{\arabic{footnote}}
        \setcounter{footnote}{0}}  

\numberbysection \hybrid

\newcommand{\tr}{{\rm tr}}

\newcommand{\al}{\alpha}
\newcommand{\be}{\beta}
\newcommand{\ga}{\gamma}

\newcommand{\vth}{\vartheta}

\newcommand{\bfq}{{\bf{q}}}
\newcommand{\bfx}{{\bf{x}}}

\def\beq{\begin{equation}}
\def\eeq{\end{equation}}
\def\p{\partial}

\newtheorem{theor}{Theorem}

\def\res{\mathop{\hbox{Res}}\limits}

\begin{document}

\setcounter{page}{1}

\date{}
\date{}
\vspace{50mm}

\begin{flushright}
 ITEP-TH-27/13\\
\end{flushright}
\vspace{3mm}


\begin{center}
{\LARGE{Rational Top and its Classical R-matrix } }
\\
\vspace{8mm} {\large { G. Aminov}\,$^{\sharp\, \natural}$\ \ \ { S.
Arthamonov}\,$^{\flat\, \sharp}$\ \ \ {A. Smirnov}\,$^{\dag\,
\sharp}$\ \
\ {A. Zotov}\,$^{\diamondsuit\, \sharp\, \natural}$}\\
 \vspace{5mm}
 \vspace{3mm} $^\sharp$ - {\small{\em 
 ITEP, B. Cheremushkinskaya str. 25, 117218,  Moscow, Russia}}\\
 \vspace{2mm}$^\natural$ - {\small{\em Moscow Institute of Physics and Technology, Inststitutskii per.  9, 141700,\\ Dolgoprudny,
 Moscow region, Russia}}\\
 \vspace{2mm}
  $^\flat$ - {\small{\em Rutgers University, Department of Mathematics, 110 Frelinghuysen
  Road\\
Piscataway, NJ 08854-8019, USA}}\\
 \vspace{2mm}
  $^\dag$ - {\small{\em Columbia University, Department of Mathematics,  MC
  4406,
2990 Broadway\\
New York, NY 10027, USA}}\\
 \vspace{2mm} $^\diamondsuit$ - {\small{\em Steklov Mathematical
Institute,  RAS, Gubkina str. 8, 119991, Moscow,  Russia}}
\\ \ \\  \ \\
%
\end{center}
 \begin{abstract}
We construct a rational integrable system (the rational top) on a
coadjoint orbit of ${\rm SL}_N$ Lie group. It is described by the
Lax operator with spectral parameter and classical non-dynamical
skew-symmetric $r$-matrix. In the case of the orbit of minimal
dimension the model is gauge equivalent to the rational
Calogero-Moser (CM) system. To obtain the results we represent  the
Lax operator of the CM model in two different factorized forms --
without spectral parameter (related to the spinless case) and
another one with the spectral parameter. The latter gives rise to
the rational top while the first one is  related to generalized
Cremmer-Gervais $r$-matrices. The gauge transformation relating the
rational top and CM model provides the classical rational version of
the IRF-Vertex correspondence. From the geometrical point of view it
describes the modification of ${\rm SL}(N,\mathbb C)$-bundles over
degenerated elliptic curve. In view of the Symplectic Hecke
Correspondence the rational top is related to the rational spin CM
model. Possible applications and generalizations of the suggested
construction are discussed. In particular, the obtained $r$-matrix
defines a class of KZB equations.
 \end{abstract}



\begin{center}\footnotesize{{\rm E-mails:} {\rm
aminov@itep.ru}; {\rm artamonov@itep.ru}; {\rm
asmirnov@math.columbia.edu}; {\rm
 zotov@mi.ras.ru}}\end{center}


\renewcommand{\tableofcontents}{{\normalsize {\bf Contents}}}

\small{





\setcounter{section}{1}
\subsubsection*{1 IRF-Vertex and/or Symplectic Hecke correspondences}
\setcounter{equation}{0}

We propose the classical rational version of the IRF-Vertex (or
Face-Vertex) correspondence. In the elliptic and trigonometric cases
it was found in \cite{Hasegawa12} and \cite{Zabrodin1} respectively.
In those papers authors suggested explicit formulae for the twists
relating non-dynamical \cite{Baxter,Baxter2,CG} and dynamical
\cite{ndyn,ABB} quantum $R$-matrices (with spectral parameter) by
representing $L$-operators in the factorized forms
\cite{Sklyanin,factor2,factor}. The phenomenon of the IRF-Vertex
correspondence \cite{Baxter,irf-vertex} was observed many times at
different levels
\cite{AlF,KuznSkl,AlF2,Vakulenko,Chen,Feher1,Hodges1}.
We are motivated by geometric approach to integrable systems
\cite{Krich2,LOZ1}, where the Lax operators with spectral parameter
are considered as sections of some bundles over algebraic curves. In
this approach the IRF-Vertex twists are treated as modifications of
bundles (or Hecke transformations) \cite{Vakulenko}. The latter act
on the Lax operator as special gauge transformations degenerated at
some point and give rise to the  Symplectic Hecke Correspondence
\cite{LOZ1}. In this way one can describe sets of integrable systems
with gauge equivalent Lax representations corresponding to different
characteristic classes of bundles. In addition to both sides of the
IRF-Vertex correspondence this geometric description provides
intermediate integrable models and related $R$-matrices
\cite{LOSZ1,LOSZ4}.
In this paper we deal with the Hecke operator given by the following
matrix:
 \beq\label{v00035}
 \begin{array}{c}
 \displaystyle{
 \Xi_{ij}(z,{\bf {\bar q}})=(z+{\bar q}_j)^{\varrho(i)}\,,\ \
 i,j=1...N\,, \ \ \sum\limits_{k=1}^N{\bar q}_k=0
 }
 \end{array}
 \eeq
(see (\ref{v0003})-(\ref{v922})). It is the rational analogue of the
Hasegawa's elliptic twist \cite{Hasegawa12}. On the other hand it is
the modification of bundles over degenerated elliptic curve. Some
details of underlying geometric construction will be given in
\cite{AASZ3}.

The paper is organized as follows. In the next paragraph we give
explicit formula for non-dynamical $r$-matrix with spectral
parameter. Next, we describe the underlying integrable system - the
rational top (paragraph 3). It is defined on a coadjoint orbit of
${\rm SL}_N$. Then we show that in the case of the orbit of minimal
dimension the defined model is gauge equivalent to CM one (paragraph
4). The result is achieved by presenting the Lax operator in the
factorized form using (\ref{v00035}). Finally we give precise
relation between non-dynamical $r$-matrix  and that of CM model
(paragraph 5). Possible applications and generalizations are
discussed in paragraph 6. In particular, the obtained $r$-matrix
defines a class of KZB equations. Some useful formulae and details
of proofs are given in appendices.

\setcounter{section}{2}
\subsubsection*{2 Classical rational $r$-matrix}
\setcounter{equation}{0}

The purpose of the paper is to describe at classical level
the rational version of the IRF-Vertex correspondence obtained in
\cite{Hasegawa12}.
While the IRF side goes with the spin extension \cite{spin} of ${\rm
sl}_N$ Calogero-Moser model (CM) \cite{Calogero}, the Vertex side
corresponds to the rational top\footnote{The title is analogues to
the elliptic top \cite{LOZ1} which appears from the Belavin-Drinfeld
elliptic $r$-matrix \cite{BD}. A naive rational limit from the
latter gives $r(z,w)=P_{12}/(z-w)$, where $P_{12}$ is the
permutation operator.}. We show that the top is described by the
following non-dynamical $r$-matrix:
 \beq\label{v2002}
 {r}^{\hbox{\tiny{top}}}(z,w)={r}^{\hbox{\tiny{top}}}(x)=\frac{1}{Nx}\sum\limits_{i,j=1}^N E_{ij}\,\otimes
 \eeq
 $$
 \begin{array}{c}
 \displaystyle{
 \Big[\sum\limits_{\ga=0}^{\varrho(i)} x^\gamma \left(\!\begin{array}{c} \varrho(i) \\ \gamma
 \end{array}\!\right)
 E_{\varrho^{-1}(\varrho(i)-\ga),j}-\sum\limits_{\ga=0}^{\varrho(i)} x^{\ga+N-j+1}\,
 (-1)^{\varrho(j)+N}(N\!-\!j)\left(\!\begin{array}{c} \varrho(i) \\ \gamma
 \end{array}\!\right) \left(\!\begin{array}{c} N \\ j\!-\!1
 \end{array}\!\right) E_{\varrho^{-1}(\varrho(i)-\ga),\,N}
 }
 \\
  \displaystyle{
-N\sum\limits_{\ga=0}^{\varrho(i)} \sum\limits_{s=1}^{N-j}
\delta_{\varrho(i)-j+1\leq s+\ga} x^{s+\ga}\,
(-1)^{\varrho(j)+s+j-1}\left(\!\begin{array}{c} \varrho(i) \\
\gamma
 \end{array}\!\right) \left(\!\begin{array}{c} s\!+\!j\!-\!2 \\ j\!-\!1
 \end{array}\!\right)\sum\limits_{c=0}^{N-s-j+1} E_{\varrho^{-1}(\varrho(i)\!-\!\ga\!+\!c),\,\varrho^{-1}(s\!+\!j\!+\!c\!-\!1)}
 }
  \\
  \displaystyle{
-N\sum\limits_{\ga=0}^{\varrho(i)} \sum\limits_{s=1}^{N-j}
\delta_{\varrho(i)-j+1> s+\ga} x^{s+\ga}\,
(-1)^{\varrho(j)+s+j-2}\left(\!\begin{array}{c} \varrho(i) \\
\gamma
 \end{array}\!\right) \left(\!\begin{array}{c} s\!+\!j\!-\!2 \\ j\!-\!1
 \end{array}\!\right)\sum\limits_{c=0}^{s+j-2} E_{\varrho^{-1}(\varrho(i)\!-\!\ga\!-\!c\!-1),\,\varrho^{-1}(s\!+\!j\!-\!c\!-2)}
 }
 \\
\displaystyle{ +\!\sum\limits_{\ga=0}^{\varrho(i)}
\sum\limits_{s=1}^{N-j}
x^{s+\ga}\, (-1)^{\varrho(j)+s+j-1} \left(\!\begin{array}{c} \varrho(i) \\
\gamma
 \end{array}\!\right) \left(\!\begin{array}{c} \!s\!+\!j\!-\!1\! \\ \!j\!-\!1\!
 \end{array}\!\right) E_{\varrho^{-1}(\varrho(i)-\ga),\,\varrho^{-1}(s+j-1)}\!-\!\frac{\delta_{i,j}}{N}
\sum\limits_{k=1}^N\sum\limits_{c=0}^{N\!-\!k\!-2}E_{k+c,\,k+c+1}\Big]
 }
 \end{array}
 $$
where $x=z-w$, $\{E_{ij}\}$ is the standard basis in ${\rm gl}_N$:
$\left(E_{ij}\right)_{ab}=\delta_{ia}\delta_{jb}$, function
$\varrho$ is defined in (\ref{v922})
 and all arguments of $\varrho^{-1}$ in (\ref{v2002}) are
not equal to $N-1$ (the corresponding values of indices are skipped
in the sums).

In what follows we use $\delta$ as the standard Kronecker delta and
in the form: $\delta(A)=1$ if $A$ is true, $\delta(A)=0$ if $A$ is
false, where $A$ is a restriction on indices values.

For $N=2$ and $N=3$ (\ref{v2002}) takes the form:
 \beq\label{v209}
{\rm sl}(2,{\mathbb C}):\ \ \ \ \
{r}^{\hbox{\tiny{top}}}_{12}(z,w)=\frac{1}{2}\,
 \left(\begin{array}{cccc}
1/x & 0 & 0 & 0\\ -x & 0 & 1/x & 0\\ -x & 1/x & 0 & 0\\ -x^3 & x & x
& 1/x
 \end{array}
 \right)-\frac{1_{4\times 4}}{4x}\,,\ \ \ \ x=z-w\,.
 \eeq
 \beq\label{v210}
{\rm sl}(3,{\mathbb C}):\ \
{r}^{\hbox{\tiny{top}}}_{12}(z,w)=\frac{1}{3}\,
 \left(\begin{array}{ccccccccc}
1/x & 0 & 0 & 0 & 0 & 0 & 0 & 0 & 0\\ 1 & 0 & 0 & 1/x & 0 & 0 & 0 & 0 & 0\\
2 x^2 & -3 x & 0 & -3
x & 3 & 0 & 1/x & 0 & 0\\ -1 & 1/x & 0 & 0 & 0 & 0 & 0 & 0 & 0\\ 2 x & 0 & 0 & 0 & 1/x & 0 & 0 & 0 & 0\\
2 x^3 & -x^2 & 1 & -3 x^2 & 0 & 0 & 1 & 1/x & 0\\ -2 x^2 & -3 x &
1/x & -3 x & -3 & 0 & 0 & 0 & 0\\ 2 x^3 & 3 x^2 & -1 & x^2 & 0 & 1/x & -1 & 0 & 0\\
2 x^5 & 3 x^4 & -x^2 & -3 x^4 & -6 x^3 & 3 x & x^2 & 3 x & 1/x
 \end{array}
 \right)-\frac{1_{9\times 9}}{9x}\,,
 \eeq
where $x=z-w$.  The following statement holds:
\begin{theor}\label{theor1}

\

 \noindent {\bf 1.} The $r$-matrix (\ref{v2002}) is skew-symmetric
and satisfies the classical Yang-Baxter equation
 \beq\label{v242}
 \begin{array}{c}
 \displaystyle{
[{r}^{\hbox{\tiny{top}}}_{12}(z,w),{
r}^{\hbox{\tiny{top}}}_{13}(z,u)]+[{r}^{\hbox{\tiny{top}}}_{12}(z,w),{
r}^{\hbox{\tiny{top}}}_{23}(w,u)]+[{r}^{\hbox{\tiny{top}}}_{13}(z,u),{
r}^{\hbox{\tiny{top}}}_{23}(w,u)]=0\,.
 }
 \end{array}
 \eeq

 \noindent {\bf 2.} The $r$-matrix (\ref{v2002}) is gauge equivalent to the $r$-matrix of rational ${\rm sl}_N$ Calogero-Moser
 model with the gauge transformation
 \beq\label{v0002}
 \begin{array}{c}
 \displaystyle{
 g=\Xi(z,{\bar{\bf q}})\,D^{-1}({\bar{\bf q}})\,,
 }
 \end{array}
 \eeq
 where $\Xi(z,{\bar{\bf q}})$ is given by (\ref{v00035}),
  \beq\label{v00052}
 \begin{array}{c}
 \displaystyle{
 D_{ij}(\bar\bfq)=D_{ij}(\bfq)=\delta_{ij}\prod\limits_{k\neq
 i}({\bar q}_i-{\bar q}_k)
 }
 \end{array}
 \eeq
 and
 ${\bar{\bf q}}=({\bar q}_1,...,{\bar q}_N)$ is the set of
 ${\rm sl}_N$ CM particles coordinates, $\bar q_j=q_j-\frac{1}{N}\sum\limits_{k=1}^N
 q_k$.

\end{theor}
The proof is direct. In fact, the first part follows from the second
one, and the second is precisely described in paragraph 5. The
skew-symmetry $r_{12}(z,w)=-r_{21}(w,z)$ (see (\ref{v3114})) can be
verified separately.

Presumably, (\ref{v2002}) is some non-trivial limit from the
Belavin-Drinfeld elliptic $r$-matrix.
The case $N=2$ (\ref{v209}) was found in \cite{Burban1} by
considering bundles over degenerated (cuspidal) elliptic curves.
Independently, the same answer was also obtained at the level of Lax
operators \cite{Smirnov1} (see also \cite{Smirnov2,Smirnov3}) using
special limiting procedure starting from the elliptic case. Let us
mention that in \cite{Burban2} the non-dynamical ${\rm sl}_N$
$r$-matrix was obtained in a different form
$r(z,w)=\frac{P_{12}}{z-w}+z\,r^1+w\,r^2$
 using approach of \cite{Stolin}, where $P_{12}$ is the permutation operator while $r^1$ and $r^2$ are some non-trivial constants.
 From the results of
\cite{BD2} it is natural to expect that the answers (\ref{v2002})
and the one from \cite{Burban2} are gauge equivalent. However, this
question deserves further elucidation.

The non-dynamical form of CM $r$-matrix was studied in \cite{Feher1}
(see also \cite{Hodges1,Forger,Og}) in the cases without spectral
parameters (including the rational one). Authors of \cite{Feher1}
used gauge
 transformation (\ref{v0002}) with $\Xi(z,{\bar{\bf q}})$ replaced
 by the Vandermonde matrix (it was originally found in \cite{FeherBalog})
 \beq\label{v00036}
 \begin{array}{c}
 \displaystyle{
 V_{ij}({\bf { q}})=q_j^{\,i-1}\,,\ \
 i,j=1...N\,.
 }
 \end{array}
 \eeq
It was shown that the spinless CM $r$-matrix without spectral
parameters is related to the Cremmer-Gervais one \cite{CG}.
We also discuss this case in
paragraph 4.

\setcounter{section}{3}
\subsubsection*{3 Rational top}
\setcounter{equation}{0}

In this paper we use  ${\rm gl}^*_N$-variables $S_{ji}$ dual to
generators $E_{ij}$ of Lie algebra ${\rm gl}_N$:
$\left(E_{ij}\right)_{ab}=\delta_{ia}\delta_{bj}$. Then the ${\rm
sl}^*_N$-variables are naturally defined as
 \beq\label{v3525}
{\bar S}_{ij}=S_{ij}-\frac{1}{N}\,\delta_{ij}\,\tr S\in {\rm
sl}^*_N\,,\ \ \ S=\sum\limits_{i,j}E_{ij}S_{ij}\in {\rm gl}_N\,,
 \eeq
i.e. we use "bar" in ${\rm sl}_N$ case. In the same manner  this
notation is used for canonical CM variables (\ref{v02157})
 \beq\label{v35258}
\bar q_j=q_j-\frac{1}{N}\sum\limits_{k=1}^N
 q_k\,,\ \ \ \bar p_j=p_j-\frac{1}{N}\sum\limits_{k=1}^N
 p_k\,.
 \eeq
Let us start with the main statement.
\begin{theor}\label{theor2}

\

\noindent The $r$-matrix (\ref{v2002}) defines the classical
integrable system (the rational
 top) on a coadjoint orbit of ${\rm SL}(N, {\mathbb C})$ Lie group.
 The phase space is parameterized by ${\bar S}\in{\rm sl}^*(N,{\mathbb C})$ with the
 Poisson-Lie structure
 \beq\label{v305}
 \begin{array}{c}
 \displaystyle{
 \{{\bar S}_{ij},{\bar S}_{kl}\}=\delta_{kj}{\bar S}_{il}-\delta_{il}{\bar S}_{kj}
 }
 \end{array}
 \eeq
and fixed eigenvalues of $\bar S$. The ${\rm sl}_N$-valued Lax
matrix
 \beq\label{v2001}
 \begin{array}{c}
 \displaystyle{
{\bar
L}_1^{\hbox{\tiny{top}}}(z)=\tr_2\big({r}^{\hbox{\tiny{top}}}_{12}(z){\bar
S}_2\big)
 }
 \end{array}
 \eeq
obeys
 \beq\label{v20014}
 \begin{array}{c}
 \displaystyle{
\{ {\bar L}_1^{\hbox{\tiny{top}}}(z), {\bar
L}_2^{\hbox{\tiny{top}}}(w) \}=[{r}^{\hbox{\tiny{top}}}_{12}(z-w),
{\bar L}_1^{\hbox{\tiny{top}}}(z)+ {\bar
L}_2^{\hbox{\tiny{top}}}(w)]
 }
 \end{array}
 \eeq
and provides the set of integrals of motion (with respect to the
Poisson structure (\ref{v305})) as coefficients of the spectral
curve $\det(\lambda-{\bar L}^{\hbox{\tiny{top}}}(z))=0$.
\end{theor}
The proof of the theorem follows from the Yang-Baxter equation
(\ref{v242}).

Explicit form of the Lax matrix (\ref{v2001}) follows from
(\ref{v2002}) and (\ref{v2001}):
 \beq\label{v352}
 {\bar L}^{\hbox{\tiny{top}}}_{ij}(z)=\frac{1}{Nz}\times
 \eeq
 $$
\begin{array}{c}
 \displaystyle{
  \Big[\sum\limits_{\ga=0}^{\varrho(i)} z^\gamma \left(\!\begin{array}{c} \varrho(i) \\ \gamma
 \end{array}\!\right)
 S_{\varrho^{-1}(\varrho(i)-\ga),j}-\sum\limits_{\ga=0}^{\varrho(i)} z^{\ga+N-j+1}\,
 (-1)^{\varrho(j)+N}(N\!-\!j)\left(\!\begin{array}{c} \varrho(i) \\ \gamma
 \end{array}\!\right) \left(\!\begin{array}{c} N \\ j\!-\!1
 \end{array}\!\right) S_{\varrho^{-1}(\varrho(i)-\ga),\,N}
 }
 \\
  \displaystyle{
-N\sum\limits_{\ga=0}^{\varrho(i)} \sum\limits_{s=1}^{N-j}
\delta_{\varrho(i)-j+1\leq s+\ga} z^{s+\ga}\,
(-1)^{\varrho(j)+s+j-1}\left(\!\begin{array}{c} \varrho(i) \\
\gamma
 \end{array}\!\right) \left(\!\begin{array}{c} s\!+\!j\!-\!2 \\ j\!-\!1
 \end{array}\!\right)\sum\limits_{c=0}^{N-s-j+1} S_{\varrho^{-1}(\varrho(i)\!-\!\ga\!+\!c),\,\varrho^{-1}(s\!+\!j\!+\!c\!-\!1)}
 }
  \\
  \displaystyle{
-N\sum\limits_{\ga=0}^{\varrho(i)} \sum\limits_{s=1}^{N-j}
\delta_{\varrho(i)-j+1> s+\ga} z^{s+\ga}\,
(-1)^{\varrho(j)+s+j-2}\left(\!\begin{array}{c} \varrho(i) \\
\gamma
 \end{array}\!\right) \left(\!\begin{array}{c} s\!+\!j\!-\!2 \\ j\!-\!1
 \end{array}\!\right)\sum\limits_{c=0}^{s+j-2} S_{\varrho^{-1}(\varrho(i)\!-\!\ga\!-\!c\!-1),\,\varrho^{-1}(s\!+\!j\!-\!c\!-2)}
 }
 \\
  \displaystyle{
+\!\sum\limits_{\ga=0}^{\varrho(i)} \sum\limits_{s=1}^{N-j}
z^{s+\ga}\, (-1)^{\varrho(j)+s+j-1} \!\left(\!\begin{array}{c} \varrho(i) \\
\gamma
 \end{array}\!\right) \left(\!\begin{array}{c} s\!+\!j\!-\!1 \\ j\!-\!1
 \end{array}\!\right) S_{\varrho^{-1}(\varrho(i)-\ga),\,\varrho^{-1}(s+j-1)}\!-\!\frac{\delta_{i,j}}{N}
\sum\limits_{k=1}^N\sum\limits_{c=0}^{N\!-\!k\!-2}S_{k+c,\,k+c+1}\Big]
 }
 \end{array}
 $$
It has the following structure:
 \beq\label{v303}
 \begin{array}{c}
 \displaystyle{
 {\bar L}^{\hbox{\tiny{top}}}(z)=\frac{1}{Nz}{\bar S}+\frac{1}{N}\sum\limits_{k=0}^{2N-1}z^k
 \stackrel{k}{{\mathcal J}}(S)=\frac{1}{N}\sum\limits_{k=-1}^{2N-1}z^k
 \stackrel{k}{{\mathcal J}}(S)\,,\ \ \ \stackrel{-\!1}{{\mathcal
 J}}(S)={\bar S}\,,
 }
 \end{array}
 \eeq
 where the coefficients of the expansion are defined by linear constant operators on ${\rm gl}(N,{\mathbb
C})$:
 \beq\label{v311}
 \begin{array}{c}
 \displaystyle{
\stackrel{k}{{\mathcal
J}}(S)=\sum\limits_{i,j=1}^N\sum\limits_{m,n=1}^N
\stackrel{k}{{\mathcal J}}_{ij\,,\,mn} S_{mn}\, E_{ij}\,.
 }
 \end{array}
 \eeq
Using these notations the $r$-matrix (\ref{v2002}) takes the form:
 \beq\label{v31129}
 \begin{array}{c}
 \displaystyle{
{\bar
r}^{\hbox{\tiny{top}}}(z,w)=\frac{1}{N}\sum\limits_{i,j=1}^N\sum\limits_{m,n=1}^N
\sum\limits_{k=-1}^{2N-1} (z-w)^k \stackrel{k}{{\mathcal
J}}_{ij\,,\,mn} E_{ij}\otimes E_{nm}
 }
 \end{array}
 \eeq
The property of skew-symmetry means that
 \beq\label{v3114}
 \begin{array}{c}
 \displaystyle{
\stackrel{k}{{\mathcal
J}}_{ij\,,\,mn}=(-1)^{k+1}\stackrel{k}{{\mathcal J}}_{mn\,,\,ij}\,.
 }
 \end{array}
 \eeq
For $N=2$ and $N=3$ (\ref{v352}) takes the form:
 \beq\label{v353}
 \begin{array}{c}
 \displaystyle{
{\rm sl}(2,{\mathbb C}):\ \ \ \ \ {\bar
L}^{\hbox{\tiny{top}}}(z)=\frac{1}{2}
 \left(
 \begin{array}{cc}
 \frac{1}{2z}\left(S_{11}-S_{22}\right)-z S_{12} & \frac{1}{z} S_{12}
 \\ \ \\
\frac{1}{z} S_{21}-z(S_{11}-S_{22})-z^3S_{12} &
\frac{1}{2z}\left(S_{22}-S_{11}\right)+z S_{12}
 \end{array}
 \right)
 }
 \end{array}
 \eeq
 \beq\label{v354}
 \begin{array}{l}
 \displaystyle{
{\rm sl}(3,{\mathbb C}):\ \ \ \ \ {\bar
L}^{\hbox{\tiny{top}}}(z)=\hskip10cm
 }
 \end{array}
 \eeq
  {\footnotesize{
 $$
 \frac{1}{3}\left(
 \begin{array}{ccc}
  \frac{1}{z}\frac{2S_{11}-S_{22}-S_{33}}{3}+S_{12}-3zS_{23}+2z^2S_{13} & \frac{1}{z}S_{12}+3S_{23}-3zS_{13} & \frac{1}{z}S_{13}
 \\ \ \\
 \frac{1}{z}S_{21}+S_{33}-S_{11}+2zS_{12}-z^2S_{23}+2z^3S_{13} & \frac{1}{z}\frac{2S_{22}-S_{11}-S_{33}}{3}-3z^2S_{13} &
 \frac{1}{z}S_{23}+S_{13}
 \\ \ \\
 \frac{1}{z}S_{31}-S_{32}-3zS_{21}-z^2(2S_{11}-3S_{22}+S_{33})
 & \frac{1}{z}S_{32}-3S_{21}-3z(S_{11}-S_{33}) &  \frac{1}{z}\frac{2S_{33}-S_{11}-S_{22}}{3}
 \\
 +2z^3 S_{12}+3z^4 S_{23}+2z^5 S_{13} & +z^2 S_{12}-6z^3 S_{23}-3z^4S_{13}
 & -S_{12}+3zS_{23}+z^2S_{13}
 \end{array}
 \right)
 $$
 }}
It follows from Theorem \ref{theor2} that $\{\tr\left({\bar
L}^{\hbox{\tiny{top}}}(z)\right)^m, \tr\left({\bar
L}^{\hbox{\tiny{top}}}(w)\right)^n \}=0$. Therefore, traces of
powers of (\ref{v352}) are the generating functions of the
Hamiltonians. The coefficients behind the highest order poles are
the Casimir functions:
 \beq\label{v3540}
\tr\left({\bar
L}^{\hbox{\tiny{top}}}(z)\right)^m=\frac{1}{(Nz)^m}\,\tr {\bar
S}^m+...\,,\ \ \ m=1\,...\,N\,.
 \eeq
Let us recall standard arguments for the Liouville-Arnold
integrability of the models with the Lax matrices of type
(\ref{v303}). The phase space is a generic ${\rm SL}(N,{\mathbb
C})$-orbit. Its dimension equals $N^2-N$ because only eigenvalues of
$\bar S$ are fixed. It is easy to see that the number of
coefficients behind the nonnegative powers of $z$ in the set
(\ref{v3540}) equals $N(N+1)/2$. Indeed, expansion of (\ref{v3540})
gives
 \beq\label{v3541}
\frac{1}{m}\tr\left({\bar
L}^{\hbox{\tiny{top}}}(z)\right)^m=\frac{1}{z^m}H^{\hbox{\tiny{top}}}_{m,m}+\frac{1}{z^{m-1}}H^{\hbox{\tiny{top}}}_{m,m-1}+...+
 H^{\hbox{\tiny{top}}}_{m,0}+...\,,\
\ \ m=1\,...\,N\,.\footnote{In fact, all coefficients
$H^{\hbox{\tiny{top}}}_{m,k<0}$ behind positive powers of $z$ vanish
because ${\bar L}^{\hbox{\tiny{top}}}(z)$ is gauge equivalent to
spin CM model which Lax matrix has only simple pole in $z$ and no
positive powers. We prove this statement elsewhere. It follows from
this argument that $\stackrel{k}{{\mathcal J}}(S)$ satisfy a set of
relations like $\tr\left(\stackrel{2N-1}{{\mathcal J}}(S)
\stackrel{2N-1}{{\mathcal J}}(S)\right)=0,\dots$}
 \eeq
 The Hamiltonians $H^{\hbox{\tiny{top}}}_{m,1}=0$ are trivial due to (\ref{v3114}). Then
 the total number of the non-trivial Hamiltonians can be computed as
 $\sum\limits_{k=1}^Nk=N(N+1)/2$.
Subtracting then the number of the Casimir functions
$H^{\hbox{\tiny{top}}}_{m,m}$ (equal to $N$) we get exactly the half
of dimension of the phase space. Equation (\ref{v20014}) guaranties
that the coefficients are the Poisson commuting Hamiltonians, i.e.
$\{H^{\hbox{\tiny{top}}}_{i,j}\,,H^{\hbox{\tiny{top}}}_{m,n}\}=0$.
The verifying that these Hamiltonians are independent is more
complicated task. We hope to prove it in our future publications.

The coefficients $\stackrel{k}{{\mathcal J}}$ (\ref{v311}) can be
found from explicit formula (\ref{v352}). This allows to compute
Hamiltonians. For example, the quadratic Hamiltonian is generated by
$\tr\left({\bar L}^{\hbox{\tiny{top}}}(z)\right)^2$. It can be
written as
 \beq\label{v3547}
H_{2,0}^{\hbox{\tiny{top}}}=\frac{1}{2}\,\tr\left(\stackrel{0}{{\mathcal
J}}(S) \stackrel{0}{{\mathcal J}}(S)\right)+\tr\left({\bar
S}\stackrel{1}{{\mathcal J}}(S)\right)\,.
 \eeq
 For $N=2$
  \beq\label{v3548}
{\rm sl}(2,{\mathbb C}):\ \ \ \ \ H_{2,0}^{\hbox{\tiny{top}}}=-{\bar
S}_{11}S_{12}\,.
 \eeq
 For $N=3$
   \beq\label{v3549}
{\rm sl}(3,{\mathbb C}):\ \ \ \ \
H_{2,0}^{\hbox{\tiny{top}}}=\frac{1}{3}S_{12}^2-S_{13}S_{21}+S_{23}({\bar
S}_{33}-{\bar S}_{11})\,.
 \eeq
 The dimension of generic ${\rm SL}_3$-orbit equals 6. Hence, we
 need two more Hamiltonians. They are
   \beq\label{v3552}
   \begin{array}{c}
   \displaystyle{
H_{3,2}^{\hbox{\tiny{top}}}=\frac{1}{27}\Big(S_{12}^2S_{21}+3S_{21}S_{23}{\bar
S}_{22}-3S_{13}S_{32}{\bar S}_{11}-5S_{12}{\bar S}_{11}{\bar
S}_{22}+S_{12}S_{13}S_{31}-2S_{32}S_{12}S_{23}}\\
\ \\ \displaystyle{+6{\bar S}_{11}S_{23}S_{21}-2S_{12}{\bar
S}_{22}^2-3S_{13}S_{21}^2+3S_{31}S_{23}^2-2S_{12}{\bar
S}_{11}^2\Big)\,, }
 \\ \ \\
H_{3,0}^{\hbox{\tiny{top}}}={\bar
S}_{11}^2S_{13}-\frac{1}{3}S_{12}S_{13}S_{21}-S_{21}S_{23}^2+\frac{1}{3}S_{23}S_{12}{\bar
S}_{11}+\frac{2}{3}S_{23}S_{12}{\bar
S}_{22}+\frac{2}{27}S_{12}^3+S_{13}{\bar S}_{11}{\bar S}_{22}\,.
 \end{array}
 \eeq
We will describe some details of equations of motion and Lax pairs
($M$-operators) in \cite{AASZ3}.

\setcounter{section}{4}
\subsubsection*{4 Factorized L-operators and Calogero-Moser model}
\setcounter{equation}{0}

\begin{theor}\label{theor3}

\

\noindent In the case when the coadjoint orbit is of
 minimal dimension
 \beq\label{v20018}
 \begin{array}{c}
 \displaystyle{
{\bar{\mathcal O}}^{\hbox{\tiny{min}}}:\ \ \hbox{Spec}({\bar
S})=(\nu,...,\nu,-(N-1)\nu)\,,\ \ \hbox{dim}\,{\bar{\mathcal
O}}^{\hbox{\tiny{min}}}=2N-2
 }
 \end{array}
 \eeq
the Lax operator (\ref{v2001}) is gauge equivalent to the Lax
operator of the rational Calogero-Moser model with the coupling
constant $\nu$. The gauge transformation is given by ${\rm
SL}_N$-analogue of (\ref{v0002})
 \beq\label{v00026}
 \begin{array}{c}
 \displaystyle{
 {\bar g}={\bar\Xi}(z,{\bar{\bf q}})\,D^{-1}({\bar{\bf q}})\,,\ \ \ {\bar\Xi}(z,{\bar{\bf
 q}})={\Xi}(z,{\bar{\bf q}})\frac{1}{\left(\det{\Xi}(z,{\bar{\bf q}})\right)^{1/N}}\,.
 }
 \end{array}
 \eeq
It provides the following change of variables for the case
(\ref{v20018}):
 \beq\label{v304}
 \begin{array}{c}
 \displaystyle{
{\bar S}_{ij}({\bar{\bf p}}, {\bar{\bf
q}},\nu)={N}\res\limits_{z=0}{\bar
L}^{\hbox{\tiny{top}}}_{ij}(z)=(-1)^{\varrho(j)}\,\sum_{m=1}^{N}\,\frac{{\bar
q}_{m}^{\,\varrho(i)} {\bar p}_{m} - \nu\, \partial_{{\bar q}_{m}}
{\bar q}_{m}^{\,\varrho(i)}}{ \prod\limits_{k\neq m}^{\,} ({\bar
q}_{m}-{\bar q}_{k})}\,\sigma_{\varrho(j)}({\bar
\bfq})+\nu\,\delta_{ij}\,,
 }
 \end{array}
 \eeq
where $\sigma_k({\bar \bfq})$ are the elementary symmetric functions
(\ref{v925})-(\ref{v9263}).
\end{theor}
{\bf Minimal orbit.} Again, it is convenient to start with ${\rm
gl}_N$ case. Consider the coadjoint orbit of ${\rm GL}_N$ of minimal
dimension, i.e. let
  \beq\label{v20019}
 \begin{array}{c}
 \displaystyle{
{{\mathcal O}}^{\hbox{\tiny{min}}}:\ \ \hbox{Spec}({
S})=(0,...,0,-N\nu)\,,\ \ \hbox{dim}\,{{\mathcal
O}}^{\hbox{\tiny{min}}}=2N-2
 }
 \end{array}
 \eeq
It means that $S$ is represented as a product of vector by covector
 \beq\label{v307}
 \begin{array}{c}
 \displaystyle{
S={\bf\alpha}^T\times {\bf\beta}\,\ \ \hbox{or}\ \
S_{ij}=\al_i\,\beta_j\,.
 }
 \end{array}
 \eeq
The Poisson brackets
 \beq\label{v3056}
 \begin{array}{c}
 \displaystyle{
 \{S_{ij},S_{kl}\}=\delta_{kj}S_{il}-\delta_{il}S_{kj}\,.
 }
 \end{array}
 \eeq
are realized via the Poisson brackets between components of $\al$
and $\be$. They are given by bivector $\{\al_i,\beta_j\}$ (while
$\{\al_i,\al_j\}=\{\be_i,\beta_j\}=0$) which is $N\!\times\!N$
matrix
 \beq\label{v309}
 \parallel \{\al_i,\beta_j\} \parallel=\left(\begin{array}{cccccc}
-1 & 0 & 0 & ...& 0 & 0
\\
0 & -1 & 0 & ...& 0 & 0
\\
\vdots & \vdots & \vdots & \vdots & \vdots & \vdots
\\
0 & 0 & 0 & ...& -1 & 0
\\
\beta_1 & \beta_2 & ... & \beta_{N-2} & \beta_{N-1} & 0
 \end{array}\right)
 \eeq
It differs from the (anti)canonical one $-1_{N\!\times\!N}$  by
non-trivial row $\{\al_N,\beta_j\}$. The latter comes from the
Poisson (Dirac) reduction generated by constrains
$\sum\limits_{k=1}^N\al_k\beta_k=-\nu N$ and $\beta_N=1$. The
reduction reduces the rank of the Poisson structure by one and makes
it equal to the half of dimension of the phase space (\ref{v20019}).
The similar description for ${\rm SL}_N$-orbit appears simply from
(\ref{v3525}).

The proof of Theorem \ref{theor3} is based on the factorized form of
the Lax matrix of Calogero-Moser model (CM). 

\noindent {\bf  Rational CM without spectral parameter.} Let $p_i$
and $q_i$, $i=1...N$ are canonically conjugated momenta and
coordinates of CM particles
 \beq\label{v02156}
 \begin{array}{c}
 \displaystyle{
 \{p_i,q_j\}=\delta_{ij}\,,\ \ \ \{p_i,p_j\}= \{q_i,q_j\}=0
 }
 \end{array}
 \eeq
and
 \beq\label{v02157}
 \begin{array}{c}
 \displaystyle{
 {\bar q}_i=q_i-\frac{1}{N}\sum\limits_{k=1}^N q_k\,,\ \ {\bar p}_i=p_i-\frac{1}{N}\sum\limits_{k=1}^N
 p_k\,.
 }
 \end{array}
 \eeq
 are those in the center of mass frame. The Lax matrix in the ${\rm
 sl}_N$ case with the coupling constant $\nu$ can be written as follows:
 \beq\label{v001}
 \begin{array}{c}
 \displaystyle{
 L^{\hbox{\tiny{CM}}}_{ij}=\delta_{ij}\left({\bar p}_i-\nu\sum\limits_{k\neq
 i}^N\frac{1}{q_i-q_k}\right)+(1-\delta_{ij})\frac{\nu}{q_i-q_j}
 }
 \end{array}
 \eeq
It differs from the custom one by the canonical map
 \beq\label{v007}
 \begin{array}{c}
 \displaystyle{
p_i\ \rightarrow\ p_i+a\sum\limits_{k\neq
 i}^N\frac{1}{q_i-q_k}=p_i+\p_j\log\det V^a
 }
 \end{array}
 \eeq
with the constant $a=-\nu$. It can be verified directly (see
\cite{AASZ3}) that the Lax matrix (\ref{v001}) is represented in the
following form:
 \beq\label{v002}
 \begin{array}{c}
 \displaystyle{
 L^{\hbox{\tiny{CM}}}={\bar P}-\nu\, g_0^{-1}\p_z g_0\,,\ \
 g_0=g_0(z,{\bar\bfq})=VD^{-1}
 }
 \end{array}
 \eeq
where
 \beq\label{v003}
 \begin{array}{c}
 \displaystyle{
 {\bar P}_{ij}=\delta_{ij}{\bar p}_i\,,\ \ \
 D_{ij}=\delta_{ij}\prod\limits_{k\neq
 i}(q_i-q_k)\,,}
 \end{array}
 \eeq
 \beq\label{v0031}
 \begin{array}{c}
 \displaystyle{ V_{ij}=V_{ij}(z,\bar\bfq)=(z+{\bar q}_j)^{i-1} }\,.
 \end{array}
 \eeq
Equivalently,
 \beq\label{v004}
 \begin{array}{c}
 \displaystyle{
 L^{\hbox{\tiny{CM}}}={\bar P}-\nu DV^{-1} C_0 V
 D^{-1}\,,
 }
 \end{array}
 \eeq
 \beq\label{v005}
 \begin{array}{c}
(C_{0})_{ij}= \left\{\begin{array}{l} j\,,\ \ \ i=j+1\,,\
i=2\,,...\,,N,\\  0\,,\ \ \  {otherwise}
\end{array}\right.
\end{array}
  \eeq
since $\p_z V=C_0 V$.    In spite of the fact that matrix $V$
(\ref{v0031}) depends on $z$ explicitly, the Lax matrix (\ref{v001})
is independent of $z$. It happens because
 \beq\label{v00563}
 \begin{array}{c}
 V(z,\bar\bfq)=B(z)\,V(0,\bar\bfq)\,,\ \ \ B_{ij}(z)=\delta_{i\geq j}
 \left(\!\begin{array}{c} i-1 \\ j-1 \end{array}\!\right)
 z^{i-j}\,,\ \ \ B(z)=e^{zC_0}\,.
\end{array}
  \eeq
Therefore,
 \beq\label{v00564}
 \begin{array}{c}
B^{-1}(z)=B(-z)\,,\ \ \ B^{-1}(z)\p_zB(z)=C_0\,.
\end{array}
  \eeq

\noindent {\bf  Rational CM with spectral parameter.} In this case
the Lax matrix is given by naive rational limit from the elliptic
one suggested in \cite{Krich1}. In ${\rm gl}_N$ case it can be
written as
 \beq\label{v021}
 \begin{array}{c}
 \displaystyle{
 L^{\hbox{\tiny{CM}}}_{ij}(z)=\delta_{ij}\left(p_i-\nu\sum\limits_{k\neq
 i}^N\frac{1}{q_i-q_k}\right)+(1-\delta_{ij})\frac{\nu}{q_i-q_j}-\frac{1}{N}\frac{\nu}{z}=
 L^{\hbox{\tiny{CM}}}_{ij}-\frac{1}{N}\frac{\nu}{z}\,.
 }
 \end{array}
 \eeq
The diagonal part here also differs from the custom one by the
canonical map (\ref{v007}). In the ${\rm sl}_N$ case we have
 \beq\label{v02106}
 \begin{array}{c}
 \displaystyle{
 {\bar L}^{\hbox{\tiny{CM}}}_{ij}(z)=\delta_{ij}\left({\bar p}_i-\nu\sum\limits_{k\neq
 i}^N\frac{1}{q_i-q_k}\right)+(1-\delta_{ij})\left(\frac{\nu}{q_i-q_j}-\frac{1}{N}\frac{\nu}{z}\right)\,.
 }
 \end{array}
 \eeq
Similarly to (\ref{v002}) the Lax matrices are presented in the
forms:
 \beq\label{v022}
 \begin{array}{c}
 \displaystyle{
 L^{\hbox{\tiny{CM}}}(z)=P-\nu\, g^{-1}\p_z g\,,\ \ \ g=\Xi(z,\bar\bfq)
 D^{-1}(\bar\bfq)
 }
 \end{array}
 \eeq
with $D$ is from (\ref{v003}) and $\Xi$ is from (\ref{v00035}) or
(\ref{v0003}) (c.f. (\ref{v0002})). And
 \beq\label{v02247}
 \begin{array}{c}
 \displaystyle{
 {\bar L}^{\hbox{\tiny{CM}}}(z)={\bar P}-\nu\, {\bar g}^{-1}\p_z {\bar g}\,,
 }
 \end{array}
 \eeq
where ${\bar g}$ is from (\ref{v00026}).

\noindent {\bf Hecke transformation}
acts on (\ref{v02247}) in a very simple way -- by gauge
transformation (see \cite{LOZ1}):
 \beq\label{v02248}
 \begin{array}{c}
 \displaystyle{
 {\bar L}^{\hbox{\tiny{CM}}}(z)\ \longrightarrow\ {\bar g}(z)\, {\bar L}^{\hbox{\tiny{CM}}}(z)\, {\bar g}^{-1}(z)=
 {\bar g}(z)\,{\bar P}\, {\bar g}^{-1}(z)-\nu\, \p_z {\bar g}(z)\,{\bar
 g}^{-1}(z)=
 }
 \end{array}
 \eeq
 $$
={\bar \Xi}(z)\,{\bar P}\, {\bar \Xi}^{-1}(z)-\nu\, \p_z {\bar
\Xi}(z)\,{\bar
 \Xi}^{-1}(z)
 $$
The statement of the Theorem \ref{theor3} is that this matrix
coincides with (\ref{v352}) in the case (\ref{v20018}) with the
change of variables (\ref{v304}):
 \beq\label{v02249}
 \begin{array}{c}
 \displaystyle{
 \left.{\bar L}^{\hbox{\tiny{top}}}(z)\right|_{{\mathcal O}^{\hbox{\tiny{min}}}}
  \stackrel{(\ref{v304})}{=} {\bar g}(z)\, {\bar L}^{\hbox{\tiny{CM}}}(z)\, {\bar g}^{-1}(z)\,.
 }
 \end{array}
 \eeq
The proof is similar to the one of Theorem \ref{theor4} which is
given in Appendix B. Let us just mention that the residue of
(\ref{v02249}) is easily calculated via (\ref{v927}). For variables
(\ref{v307}) it gives the following "bosonization"
formulae\footnote{In quantum case one can quantize the Poisson
brackets (\ref{v307})-(\ref{v309}) as
$\al_j\rightarrow\hbar\p_{\beta_j}$.}:
 \beq\label{v308}
 \begin{array}{c}
 \displaystyle{
\al_i=\sum_{m=1}^{N}\,\frac{{\bar q}_{m}^{\,\varrho(i)} p_{m} -
\nu\,{\varrho(i)}\,
 {\bar q}_{m}^{\,\varrho(i)-1}}{
\prod\limits_{k\neq m}^{\,} ({\bar q}_{m}-{\bar q}_{k})}\,,\ \ \
\beta_j=(-1)^{\varrho(j)}\,\sigma_{\varrho(j)}({\bar q}_{1},{\bar
q}_{2},...,{\bar q}_{N})\,.
 }
 \end{array}
 \eeq
One can verify that the Poisson bivector (\ref{v309}) is reproduced
via the canonical Poisson brackets (\ref{v02156}). Then
(c.f.(\ref{v304}))
 \beq\label{v30455}
 \begin{array}{c}
 \displaystyle{
{ S}_{ij}({\bar{\bf p}}, {\bar{\bf
q}},\nu)=\al_i\be_j=(-1)^{\varrho(j)}\,\sum_{m=1}^{N}\,\frac{{\bar
q}_{m}^{\,\varrho(i)} {\bar p}_{m} - \nu\, \partial_{{\bar q}_{m}}
{\bar q}_{m}^{\,\varrho(i)}}{ \prod\limits_{k\neq m}^{\,} ({\bar
q}_{m}-{\bar q}_{k})}\,\sigma_{\varrho(j)}({\bar \bfq})\,.
 }
 \end{array}
 \eeq
The obtained { bosonization} formulae (\ref{v304}) are naturally
 generalized to the case of differential operators. Replace the momenta
 $p_m$ in (\ref{v304}) as $p_m\rightarrow\hbar\p_{q_m}$. Then the
 corresponding differential operators $\hat{S}_{ij}$ commute as
 generators of the Lie algebra ${\rm gl}_N$:
 $
 [\hat{S}_{ij},\hat{S}_{kl}]=\hbar\left(\delta_{kj}\hat{S}_{il}-\delta_{il}\hat{S}_{kj}\right)$.
 In the elliptic case when the matrix $\Xi$ is given as in
 (\ref{v0004}) this type formulae were obtained
  in  \cite{Hasegawa12}.

\noindent {\bf Cremmer-Gervais top.} Let us perform a similar gauge
transformation in the case of CM without spectral parameter
(\ref{v002})-(\ref{v0031}):
 \beq\label{v341}
 \begin{array}{c}
 \displaystyle{
L^{\hbox{\tiny{CM}}}\ \longrightarrow\
g_0(z)L^{\hbox{\tiny{CM}}}g_0^{-1}(z)=
L^{\hbox{\tiny{CG}}}(z)=V(z,\bar\bfq){\bar
P}\,V^{-1}(z,\bar\bfq)-\nu\,\p_z V(z,\bar\bfq) V^{-1}(z,\bar\bfq)\,.
 }
 \end{array}
 \eeq
Using (\ref{v00563})-(\ref{v00564}) we see that
 \beq\label{v34172}
 \begin{array}{c}
 \displaystyle{
L^{\hbox{\tiny{CG}}}(z)=V(z,\bar\bfq){\bar
P}\,V^{-1}(z,\bar\bfq)-\nu\, C_0=B(z) L^{\hbox{\tiny{CG}}}(0)
B^{-1}(z)\,,}
 \end{array}
 \eeq
 \beq\label{v34173}
 \begin{array}{c}
  \displaystyle{
L^{\hbox{\tiny{CG}}}(0)=V(0,\bar\bfq){\bar
P}\,V^{-1}(0,\bar\bfq)-\nu\, C_0\,.
 }
 \end{array}
 \eeq
It means that the spectral parameter is fictive here, and can be
gauged away by non-dynamical gauge transformation $B(z)$. The latter
reflects the fact that initial model (CM) describes only $N-1$
degrees of freedom.
It appears that the Lax matrix (\ref{v34173}) can be also written in
terms of ${\rm gl}^*_N$ variables $S_{ij}$ (\ref{v30455}).
 \begin{theor}\label{theor4}

The Lax matrix (\ref{v34173}) can be written in terms of ${\rm
gl}^*_N$-variables $S_{ab}$ (\ref{v30455}) as follows:
 \beq\label{v342}
 \begin{array}{c}
 \displaystyle{
L^{\hbox{\tiny{CG}}}_{ij}(0)=\delta_{j\geq\,
i-1}\sum\limits_{c=0}^{N-j}\Big(
S_{\varrho^{-1}(i+c-1),\,\varrho^{-1}(j+c)}
-\delta_{j,\,i-1}\frac{1}{N}\sum\limits_{l=1}^NS_{ll}\Big)
 }
\\ \ \\
  \displaystyle{
-\delta_{j<i-1}\sum\limits_{c=0}^{j-1}S_{\varrho^{-1}(i-c-2),\,\varrho^{-1}(j-c-1)}-\delta_{i,j}\frac{1}{N}
\sum\limits_{k=1}^N\sum\limits_{c=0}^{N-k-2}S_{k+c,\,k+c+1}\,,
 }
 \end{array}
 \eeq
 where $\varrho^{-1}$ is the inverse of $\varrho$ (\ref{v922}) and the sum is taken over all indices for which $\varrho^{-1}$
 is defined (i.e. when its arguments are not equal to $N-1$).
 \end{theor}
\noindent The proof is achieved by substitution  of (\ref{v30455})
into (\ref{v34173}). It is given in Appendix B.

This kind of result was obtained in \cite{Feher1} at the level of
$r$-matrices. The obtained non-dynamical $r$-matrix was shown to be
the Cremmer-Gervais one \cite{CG}. This is
 why we call the obtained model as the Cremmer-Gervais top.
 Conjugating by $B(z)$ (\ref{v00563}) we get
 $L^{\hbox{\tiny{CG}}}(z)$ which is related to Jordanian (generalized)
 $R$-matrices of Cremmer-Gervais type \cite{Gerst1,Hodges1} (see also \cite{Johnson}).
Notice that its Hamiltonians coincide with subset of rational top
Hamiltonians (\ref{v3541}):
 \beq\label{v3423}
 \begin{array}{c}
 \displaystyle{
H_k^{\hbox{\tiny{CG}}}=\frac{1}{k}\tr\left(L^{\hbox{\tiny{CG}}}\right)^k=H_{k,0}^{\hbox{\tiny{top}}}\,,\
\ \ k=1...N\,.
 }
 \end{array}
 \eeq
\noindent  This is due to (\ref{v021}).
However, this model has only $N-1$ independent Hamiltonians. It
corresponds to the minimal orbit case (\ref{v20018}), while the
rational top has more Hamiltonians and describes generic orbit.

Let us give examples (\ref{v342}) for $N=2$ and $N=3$:
 \beq\label{v349}
 \begin{array}{c}
 \displaystyle{
{\rm sl}(2,{\mathbb C}):  L^{\hbox{\tiny{CG}}}(0)=
 \left(
 \begin{array}{cc}
 0 & S_{12}
 \\ \ \\
 -{\bar S}_{11} & 0
 \end{array}
 \right)
 }
 \end{array}
 \eeq
 %
 \beq\label{v3492}
 \begin{array}{c}
 \displaystyle{
 {\rm sl}(3,{\mathbb C}):  L^{\hbox{\tiny{CG}}}(0)=
 \left(
 \begin{array}{ccc}
  \frac{2}{3}S_{12} & S_{23} & S_{13}
 \\ \ \\
 -{\bar S}_{11} & -
 \frac{1}{3}S_{12} & S_{23}
 \\ \ \\
 -S_{21} & {\bar S}_{33} & -\frac{1}{3}S_{12}
 \end{array}
 \right)
 }
 \end{array}
 \eeq

\setcounter{section}{5}
\subsubsection*{5 Rational classical IRF-Vertex
correspondence} \setcounter{equation}{0}

Let us make precise assertion of  the  second part of Theorem
\ref{theor1}. Recall that the classical $r$-matrix structure for the
rational Calogero-Moser model is given as follows \cite{r-mat,BAB2}:
 \beq\label{v200}
 \begin{array}{c}
 \displaystyle{
\{L_1^{\hbox{\tiny{CM}}}(z),L_2^{\hbox{\tiny{CM}}}(w)\}=[{r}^{\hbox{\tiny{CM}}}_{12}(z,w),L_1^{\hbox{\tiny{CM}}}(z)]-
[{r}^{\hbox{\tiny{CM}}}_{21}(w,z),L_2^{\hbox{\tiny{CM}}}(w)]\,.
 }
 \end{array}
 \eeq
 \beq\label{v201}
 \begin{array}{c}
 \displaystyle{
{r}^{\hbox{\tiny{CM}}}_{12}(z,w)=r^0_{12}(z,w)+\frac{1}{Nw}\sum\limits_{i}
E_{ii}\otimes E_{ii}+\sum\limits_{i\neq j}\left(
\frac{1}{q_i-q_j}+\frac{1}{Nw} \right)E_{ii}\otimes E_{ji}\,,
 }
 \end{array}
 \eeq
where $r^0_{12}$ is the r-matrix of the spin Calogero model:
 \beq\label{v202}
 \begin{array}{c}
 \displaystyle{
{r}^0_{12}(z,w)=\frac{1}{N}\frac{1}{z-w}\sum\limits_{i,j}
E_{ij}\otimes E_{ji}-\sum\limits_{i\neq j}
\frac{1}{q_i-q_j}E_{ij}\otimes E_{ji}\,.
 }
 \end{array}
 \eeq
Using notation (see (\ref{v02106}))
 \beq\label{v028}
 \begin{array}{c}
 \displaystyle{
 l^{\hbox{\tiny{CM}}}_{ij}(z)\equiv\left(D\Xi^{-1}\p_z\Xi D^{-1}\right)_{ij}=\delta_{ij}\left(\frac{1}{Nz}+\sum\limits_{k\neq
 i}^N\frac{1}{q_i-q_k}\right)-(1-\delta_{ij})\left(\frac{1}{q_i-q_j}-\frac{1}{Nz}\right)\,,
 }
 \end{array}
 \eeq
 we have
 \beq\label{v2022}
 \begin{array}{c}
 \displaystyle{
{r}^{\hbox{\tiny{CM}}}_{12}(z,w)=\frac{1}{N}\frac{1}{z-w}\frac{z}{w}\sum\limits_{i}
E_{ii}\otimes E_{ii}+\sum\limits_{i\neq j}
l^{\hbox{\tiny{CM}}}_{ij}(z-w) E_{ij}\otimes
E_{ji}+l^{\hbox{\tiny{CM}}}_{ij}(w) E_{jj}\otimes E_{ij}
 }
 \end{array}
 \eeq
 For our purposes we need
 \beq\label{v203}
 \begin{array}{c}
 \displaystyle{
{{\bar
r}^{\hbox{\tiny{CM}}}_{12}}(z,w)={r}^{\hbox{\tiny{CM}}}_{12}(z,w)-\frac{1}{N}\,
1\otimes l^{\hbox{\tiny{CM}}}(w)-\frac{1}{N^2}\frac{1}{z-w}1\otimes
1\,,
 }
 \end{array}
 \eeq
 which also satisfies (\ref{v200}).
The latter redefinition can be obtained by going to ${\rm
sl}_N$-valued generators together with simple dynamical twist
$r\rightarrow r+\delta r$,  $\delta r=\sum\limits_{i=1}^N \p_i\log
D\otimes E_{ii}$ with $D$ from (\ref{v003}) (see e.g. Lemma 1 in
\cite{LOSZ3}). The non-dynamical $r$-matrix appears via the gauge
transformation (\ref{v022}). It gives
 \beq\label{v206}
 \begin{array}{c}
 \displaystyle{
{r}^{\hbox{\tiny{top}}}_{12}(z,w)=g_1(z) g_2(w)\Big({{\bar
r}^{\hbox{\tiny{CM}}}_{12}}(z,w)+g_1^{-1}(z)\{g_1(z),L^{\hbox{\tiny{CM}}}_2(w)\}\Big)
g_1^{-1}(z) g_2^{-1}(w)\,,
 }
 \end{array}
 \eeq
where the second term is easily computed:
 \beq\label{v207}
 \begin{array}{c}
 \displaystyle{
g_1^{-1}(z)\{g_1(z),L^{\hbox{\tiny{CM}}}_2(w)\}
=\frac{1}{N}l^{\hbox{\tiny{CM}}}(z)\otimes 1-\frac{1}{Nz}1\otimes
1+\sum\limits_{i,j}(E_{ii}-E_{ij})\otimes E_{jj}\,
l^{\hbox{\tiny{CM}}}_{ij}(z)\,.
 }
 \end{array}
 \eeq
After cumbersome calculations one can get the resultant $r$-matrix
(\ref{v2002}). We will give the proof of the statement at quantum
level in our next paper.
%

\setcounter{section}{6}
\subsubsection*{6 Applications and remarks}
\setcounter{equation}{0}

\begin{itemize}

\item Having non-dynamical skew-symmetric $r$-matrix such that $r(z,w)=r(z-w)$
one can naturally define the Knizhnik-Zamolodchikov-Bernard (KZB)
equations \cite{KZB}. Consider tensor product $V^{\otimes n}$ of $n$
${\rm sl}_N$-modules. The $r(z,w)$-matrix acts on $V^{\otimes 2}$.
Let $r^{ab}(z_a,z_b)$ acts on $a$-th and $b$-th components of
$V^{\otimes n}$. Then the KZB equations for conformal block $\psi$
are defined as
 \beq\label{v601}
 \begin{array}{c}
 \displaystyle{
\nabla_a\psi=0\,,\ \ \ \nabla_a=\p_{z_a}+\sum\limits_{c\neq
a}r^{ac}(z_a,z_c)\,,\ \ a=1,...,n\,.
 }
 \end{array}
 \eeq
 Equations  (\ref{v601}) are compatible ($[\nabla_a,\nabla_b]=0$)
 due to the classical Yang-Baxter equation (\ref{v242}).
 The $r$-matrix (\ref{v2002}) satisfies the above mentioned conditions (it is skew-symmetric and $r(z,w)=r(z-w)$).
 Therefore, the KZB equations with $r$-matrix (\ref{v2002}) are well
 defined.

\item The KZB equations (\ref{v601}) are known to describe
quantization of the Schlesinger system \cite{Resh}. At the level of
classical mechanics it is easy to construct generalizations of the
rational top of Gaudin-Schlesinger type. Let the phase space be a
direct product of $n$ coadjoint orbits, i.e. we have the variables
${\bar S}^a$, $a=1,...,n$ with the Poisson structure be a direct sum
of (\ref{v3056}):
 \beq\label{v602}
 \begin{array}{c}
 \displaystyle{
\{{ S}^a_{ij},{ S}^b_{kl}\}=\delta^{ab}\left(\delta_{kj}{
S}_{il}^a-\delta_{il}{ S}_{kj}^a\right)\,.
 }
 \end{array}
 \eeq
The Lax operator of the Gaudin model is constructed via the one of
the rational top (\ref{v352}):
 \beq\label{v603}
 \begin{array}{c}
 \displaystyle{
L^{\hbox{\tiny{Gaudin}}}(z)=\sum\limits_{a=1}^n
L^{\hbox{\tiny{top}}}(z-z_a,{ S}^a)\,.
 }
 \end{array}
 \eeq
It also satisfies the classical exchange relations (\ref{v20014}),
and  hence defines an integrable system. It differs from the
standard rational Gaudin model and provides an alternative limit
from the elliptic Gaudin model \cite{STS}. Similarly, the
Schlesinger type model appears by  replacing (\ref{v603}) with the
connection along the curve $\p_z+L^{\hbox{\tiny{Gaudin}}}(z)$. It
leads to alternative rational limit of the elliptic Schlesinger
system \cite{CLOZ1}. We are going to describe these models in
details in our future publications. It is interesting to compare the
models with those considered in \cite{Kulish} for ${\rm sl}_2$ case
using non-dynamical $r$-matrices \cite{Stolin2}.

\item In
trigonometric case the $R$-matrix is known at quantum level from
\cite{Zabrodin1}, where the trigonometric analogue of $\Xi$
(\ref{v00035}) was found. The analogue of (\ref{v002}) is known as
well \cite{Babel3,ABB,CG}:
 \beq\label{v0004}
\begin{array}{l}
 \displaystyle{
 V_{ij}({\bf x})=e^{(i-1)x_j};\ \ D_{ij}=\delta_{ij}\prod\limits_{k\neq
 i}(e^{{\bar q}_i}-e^{{\bar q}_k});\ \
 \Xi_{ij}({\bf x})=\! \left\{\begin{array}{l} e^{(i-1)x_j}\,, i=1\,,...\,,N-1\,,\\ \ \\ e^{(N-1)x_j}
 +(-1)^N e^{-x_j}\,,
 i\!=\!N\,,
 \end{array}\right.
 }
 \end{array}
 \eeq
where $x_j={\bar q}_j+z$. In elliptic case the spectral parameter is
crucially important. The elliptic $\Xi$ was found in
\cite{Hasegawa12} at quantum level:
 \beq\label{v0005}
 \begin{array}{c}
 \displaystyle{
 \Xi_{ij}(z, {\bf q}\,|\tau) =
\theta{\left[\begin{array}{c}
\frac{i}N-\frac12\\
\frac{N}2
\end{array}
\right]}(z-N{\bar q}_j, N\tau )\,,\ \ \
D_{ij}=\delta_{ij}\prod\limits_{k\neq
 i}\vth(q_i-q_k)\,.
 }
 \end{array}
 \eeq

\item As we mentioned the rational top is actually not a rational
model but rather degenerated elliptic one \cite{Smirnov1}. It is
defined on the bundle over the curve $y^2=z^3$ \cite{Burban1}. We
hope our results may shed light on possible elliptic generalizations
of the recently investigated dualities in integrable systems
\cite{MMRZZ,GZZ}.

\item While the CM models possess  relativistic
generalization after Ruijsenaars \cite{Ruijs1}, the top-like models
have also (group) extensions of this type \cite{BDOZ}. These type of
models can be described explicitly using the factorized form of
$L$-operators. In the group case we have the following
representations without and with spectral parameter for the
Ruijsenaars-Schneider models:
 \beq\label{v0006}
 \begin{array}{c}
 \displaystyle{
 L^{\hbox{\tiny{RS}}}= g_0^{-1}(z) g_0(z-\nu\eta)\,e^{\eta P},\ \
 \ L^{\hbox{\tiny{RS}}}(z)= g^{-1}(z) g(z-\nu\eta)\,e^{\eta P}
 }
 \end{array}
 \eeq
with $g_0$ or $g$ given by (\ref{v002}), (\ref{v022}) and
(\ref{v0004})-(\ref{v0005}). Here $\eta$ is the inverse light speed.

\item It is an interesting question -- for which $g$ the expression (\ref{v0006}) gives Lax matrix of an integrable
model? A general answer is that the $\Xi(z)$ matrix should be
modification of underlying bundle. We hope to clarify this question
in \cite{AASZ3}.

\item The structure of $\bar \bfq$-argument of $\Xi(z,\bar \bfq)$ is naturally
formulated in terms of $A_{N-1}$ root system and corresponding
fundamental weights. Therefore, one can await extensions to other
roots systems of simple Lie algebras. Presumably, the $r$-matrix is
the same (in accordance with \cite{FeherBalog}) while the $r$-matrix
structure can be of reflection type.

\item Consideration of concrete examples (\ref{v209}), (\ref{v210}) of
(\ref{v2002}) show some cancellations in the complicated expression
(\ref{v2002}). We hope that the obtained answer for the classical
$r$-matrix  can be written in a compact form.

\end{itemize}


\noindent {\bf Acknowledgments.} We are grateful to A. Levin and M.
Olshanetsky for useful discussions and remarks. The work was
partially supported by RFBR grant 12-02-00594 and by leading young
scientific groups RFBR 12-01-33071 mol$\_$a$\_$ved. The work of G.A.
and A.Z. was also supported by the D. Zimin's fund "Dynasty". The
work of A.Z. was also supported by the Program of RAS "Basic
Problems of the Nonlinear Dynamics in Mathematical and Physical
Sciences".


\setcounter{section}{7}
\subsubsection*{7 Appendix A: rational modification}
\setcounter{equation}{0}

Here we collect some simple algebraic facts related to matrix
(\ref{v00035}).
We start with

\noindent {\bf Vandermonde matrix}
 \beq\label{v901}
 \begin{array}{c}
 \displaystyle{
 V_{ij}({\bf x})=x_j^{i-1}\, \ \ \ i,j=1,...,N\,.
 }
 \end{array}
 \eeq
It has the following determinant
 \beq\label{v902}
 \begin{array}{c}
 \displaystyle{
 \det V=\prod\limits_{1\geq i>j\geq N}(x_i-x_j)
 }
 \end{array}
 \eeq
and inverse
 \beq\label{v903}
 \begin{array}{c}
 \displaystyle{
{V^{-1}}_{kl}=\frac{1}{(l-1)!}\p_\mu^{l-1}\left.\prod\limits_{s\neq
k}^N\frac{\mu-x_s}{x_k-x_s}\right|_{\mu=0}\,.
 }
 \end{array}
 \eeq
The latter formula can be easily obtained by considering the set of
polynomials of degree $N-1$:
 \beq\label{v904}
f_k(\zeta)=f_k(\zeta,x_1,...,x_N)=\prod\limits_{s\neq
k}^N\frac{\zeta-x_s}{x_k-x_s}\,,\ \ \ k=1,...,N\,.
 \eeq
From  obvious property
 \beq\label{v905}
 f_k(x_j)=\delta_{kj}=\sum\limits_{l=1}^N {V^{-1}}_{kl}\, x_j^{l-1}
 \eeq
and the Taylor expansion
 \beq\label{v906}
 f_k(\zeta)=\sum\limits_{l=1}^N
\frac{1}{(l-1)!}\p_\mu^{\,l-1}\left.\prod\limits_{s\neq
k}^N\frac{\mu-x_s}{x_k-x_s}\right|_{\mu=0}\zeta^{l-1}
 \eeq
 we get (\ref{v903}) as
 \beq\label{v907}
 \begin{array}{c}
 \displaystyle{
{V^{-1}}_{kl}=\left.\frac{1}{(l-1)!}\,\p_\mu^{\,l-1}f_k(\mu)\right|_{\mu=0}\,.
 }
 \end{array}
 \eeq
 Multiplying (\ref{v905}) by $\zeta^{k-1}$ and summing up over $k$
 we come to identity:
  \beq\label{v908}
  \sum\limits_{k=1}^{N} \, x_{k}^{m-1}
 f_{k}(\zeta) = \zeta^{m-1}\,,\
 \ \ m=1,...,N\,.
 \eeq

\noindent {\bf $\Xi$ matrix} is the main object:
 \beq\label{v0003}
 \begin{array}{c}
 \displaystyle{
 \Xi(z,{\bf {\bar q}})= \left(\begin{array}{cccc}
 1 & 1 & ... & 1
 \\
 z+{\bar q}_1 & z+{\bar q}_2 & ... & z+{\bar q}_N
 \\
 \vdots & \vdots & \vdots & \vdots
 \\
 \ \ \,(z+{\bar q}_1)^{N-2} & \ \ \,(z+{\bar q}_2)^{N-2} & ... & \ \ \,(z+{\bar q}_N)^{N-2}
  \\ \ \\
  (z+{\bar q}_1)^{N} &  (z+{\bar q}_2)^{N} & ... &  (z+{\bar q}_N)^{N}
 \end{array}\right)
 }
 \end{array}
 \eeq
Using function
 \beq\label{v922}
\varrho(i)=\left\{\begin{array}{ll}
i-1 & {\rm{for}}\ \ 1\leq i\leq N-1,\\
 & \\
i & {\rm{for}}\ \ i= N.
\end{array}\right.\
 \hskip10mm
 \varrho^{-1}(i)=\left\{\begin{array}{ll}
i+1 & {\rm{for}}\ \ 0\leq i\leq N-2,\\
 & \\
i & {\rm{for}}\ \ i= N.
\end{array}\right.
 \eeq
it takes the form (\ref{v00035}). Consider
 \beq\label{v923}
 \Xi_{ij}({\bf x})=
 x_{j}^{\varrho(i)}\,.
 \eeq
The determinant equals
 \beq\label{v911}
 \begin{array}{c}
 \displaystyle{
\det{\Xi}(\bfx)=\det{V(\bfx)}\sum\limits_{s=1}^N
x_s=\left(\sum\limits_{s=1}^N x_s\right) \prod\limits_{1\geq i>j\geq
N}(x_i-x_j)\,.
 }
 \end{array}
 \eeq
In order to get (\ref{v911}) consider $N\!+\!1\!\times\!N\!+\!1$
Vandermonde matrix $\stackrel{N\!+\!1}{V}$ (depending on $N\!+\!1$
variables). The matrix $\Xi$ is obtained from
$\stackrel{N\!+\!1}{V}$ by deleting the $N$-th row and $N\!+\!1$-th
column. Therefore,
 \beq\label{v914}
 \begin{array}{c}
 \displaystyle{
 \det\Xi=(-1)^{2N+1}\,\left[\left(\stackrel{N\!+\!1}{V}\right)^{-1}\right]_{N\!+\!1,N}\det\stackrel{N\!+\!1}{V}
 }
 \end{array}
 \eeq
Substituting here (\ref{v902}) and (\ref{v903}) taken for
$N:=N\!+\!1$ we get (\ref{v911}).

 Consider the set of polynomials of degree $N$:
 \beq\label{v915} h_k(\zeta)=\left(1+\frac{\zeta-x_k}{\sum\limits_{s=1}^N
x_s}\right)\prod\limits_{s\neq
k}^N\frac{\zeta-x_s}{x_k-x_s}=\frac{\zeta-x'_k}{x_k-x'_k}\prod\limits_{s\neq
k}^N\frac{\zeta-x_s}{x_k-x_s}\,,\ \ \ x'_k=-\sum\limits_{s\neq
k}x_s\,.
 \eeq
From (\ref{v905}) we have
 \beq\label{v918}
 h_k(x_j)=\delta_{kj}\,.
 \eeq
The analogues of (\ref{v906})-(\ref{v907}) are easily obtained:
  \beq\label{v924}
 \Xi_{ij}^{-1}({\bf x}) = \left.\frac{1}{\varrho(j)!}\, \partial_{\zeta}^{\,\varrho(j)}\,h_{i}(\zeta,{\bf
 x})\right|_{\zeta=0}\,.
 \eeq
or
  \beq\label{v9242}
 \sum\limits_{j=1}^N \Xi_{ij}^{-1}({\bf x})\,\zeta^{\varrho(j)} =
 h_i(\zeta,{\bfx})\,.
 \eeq
The analogue of (\ref{v908}) reads as follows:
 \beq\label{v920}
  \sum\limits_{k=1}^{N} \, x_{k}^{m}
 h_{k}(\zeta) = \zeta^{m} +
 \frac{\prod\limits_{i=1}^{N}(\zeta-x_{i})}{\sum\limits_{i=1}^{N}\,x_{i}}\,\delta_{m,N-1}\,,\
 \ \ m=1,...,N\,.\footnote{Notice that the second term in the r.h.s. vanishes for
$m=\varrho(m')$.}
 \eeq
It is also convenient to use the elementary symmetric functions.
They appear from the expansion
 \beq\label{v925}
 \begin{array}{c}
 \displaystyle{
{\mathcal H}(\zeta,\bfx)=\prod\limits_{k=1}^{N} \,(\zeta-x_{k})=
\sum\limits_{k=0}^{N} (-1)^{k} \zeta^{k} \sigma_{k}(x_{1},...,x_{N})
 }
 \end{array}
 \eeq
or
 \beq\label{v926}
 \begin{array}{c}
 \displaystyle{
\sigma_{N-d}(\bfx)=(-1)^N\sum\limits_{1 \leq i_{1} <
i_{2}...<i_{d}\leq N} x_{i_{1}} x_{i_{2}}...x_{i_{d}}\,,\ \ \
d=0,...,N
 }
 \end{array}
 \eeq
In the same way define the set $\stackrel{k}{\sigma}_s(\bfx)$ by
 \beq\label{v9263}
 \begin{array}{c}
 \displaystyle{
-\prod\limits_{m\neq k}^{N} \,(\zeta-x_{m})=\p_k {\mathcal
H}(\zeta,\bfx)=\sum\limits_{s=0}^{N-1} (-1)^{s} \zeta^{s}
\stackrel{k}{\sigma}_s(\bfx)\,.
 }
 \end{array}
 \eeq
In this notation
  \beq\label{v9264}
 \begin{array}{c}
 \displaystyle{
V^{-1}_{kj}(\bfx)=(-1)^j\frac{\stackrel{k}{\sigma}_{j-1}(\bfx)}{\prod\limits_{s\neq
k}^{N} \,(x_k-x_{s})}\,.
 }
 \end{array}
 \eeq
Set also
$\stackrel{k}{\sigma}_N(\bfx)=\stackrel{k}{\sigma}_{-1}(\bfx)=0$.
Since
$h_k(\zeta,\bfx)=-\left(\frac{\zeta-x'_k}{x_k-x'_k}\prod\limits_{s\neq
k}^N\frac{1}{x_k-x_s}\right)\p_{x_k}{\mathcal H}(\zeta,\bfx)$, then
 \beq\label{v927}
 \begin{array}{c}
 \displaystyle{
\Xi^{-1}_{k
j}({\bfx})=(-1)^{\varrho(j)}\,\frac{1}{(\sum\limits_{s=1}^{N}
\,x_{s})\prod\limits_{s\neq k}^N
(x_{k}-x_{s})}\,\left(\stackrel{k}{\sigma}_{\varrho(j)-1}(\bfx) +
x_{k}^{\prime} \,\stackrel{k}{\sigma}_{\varrho(j)}(\bfx)  \right)
 }\,.
 \end{array}
 \eeq
The following set of identities holds:
 \beq\label{v928}
 \begin{array}{c}
x_{k} \stackrel{k}{\sigma}_{0}({\bfx})=\sigma_{0}({\bfx})\,,
 \\
 \vdots
 \\
\stackrel{k}{\sigma}_{j-1}({\bfx}) + x_{k}
\,\stackrel{k}{\sigma}_{j}({\bfx})=\sigma_{j}({\bfx})\,,\ \
\forall\,k\ \hbox{and}\ j=1...N-1\,,
 \\
 \vdots
 \\
 \stackrel{k}{\sigma}_{N-1}({\bfx})=\sigma_{N}({\bfx})\,.
 \end{array}
 \eeq
Hence,
 \beq\label{v929}
 \begin{array}{c}
 \displaystyle{
\Xi^{-1}_{k
j}({\bfx})=(-1)^{\varrho(j)}\,\frac{\sigma_{\varrho(j)}({\bfx})}{(\sum\limits_{s=1}^{N}
\,x_{s})\prod\limits_{s\neq k}^N
(x_{k}-x_{s})}-(-1)^{\varrho(j)}\,\frac{\stackrel{k}{\sigma}_{\varrho(j)}({\bfx})}{\prod\limits_{s\neq
k}^{N} (x_{k}-x_{s})} }\,,\ \ \forall\, k,j\,.
 \end{array}
 \eeq
Equations (\ref{v928}) can be considered as linear system for
expressing $\stackrel{k}{\sigma}_{i}$ in terms of $\sigma_j$. It can
be done in two ways -- using expansion in positive powers of $x_k$
 \beq\label{v931}
 \begin{array}{c}
 \displaystyle{
\stackrel{m}{\sigma}_{j}({\bfx})=\sum\limits_{c=0}^{N-j-1}
(-x_m)^{c} \sigma_{j+1+c}(\bfx)\ \ \ \hbox{or}\ \ \
\stackrel{m}{\sigma}_{j}({\bfx})=\sum\limits_{c=j+1}^N
(-x_m)^{c-j-1} \sigma_c(\bfx)
 }
 \end{array}
 \eeq
or in negative powers
 \beq\label{v9313}
 \begin{array}{c}
 \displaystyle{
\stackrel{m}{\sigma}_{j}({\bfx})=-\sum\limits_{c=0}^{j}
(-x_m)^{-1-c} \sigma_{j-c}(\bfx)\,.
 }
 \end{array}
 \eeq
In terms of the symmetric functions the inverse of
$\Xi(z,{\bar\bfq})$ (\ref{v0003}) has the following form:
 \beq\label{v934}
 \begin{array}{c}
 \displaystyle{
 \Xi^{-1}_{mj}(z,{\bf
 q})=\frac{1}{Nz}\frac{(-1)^{\varrho(j)}}{\prod\limits_{r\neq
 m}^N(q_m-q_r) }\Big(
 \sigma_{\varrho(j)}({\bar\bfq})+\sum\limits_{s=1}^{N-j} z^s\left[
 \sigma_{s+j-1}(\bar\bfq) \left(\!\begin{array}{c} s+j-1\\
 j-1\end{array}\!\right) \right.}
\\
\
\\
\displaystyle{ \left.
-N \stackrel{m}{\sigma}_{s+j-2}\!({\bar\bfq}) \left(\!\begin{array}{c} s+j-2\\
 j-1\end{array}\!\right)
 \right] -(N-j)\,z^{N-j+1} \stackrel{m}{\sigma}_{N-1}\!({\bar\bfq})  \left(\!\begin{array}{c} N\\
 j-1\end{array}\!\right) \Big)\,.
 }
 \end{array}
 \eeq

\setcounter{section}{8}
\subsubsection*{8 Appendix B: proof of Theorem \ref{theor4}}
\setcounter{equation}{0}

\noindent Let us start with the case $\nu=0$. Substitution of
$V^{-1}$ (\ref{v9264}) into (\ref{v341}) gives
 \beq\label{v345}
 \begin{array}{c}
 \displaystyle{
L^{CG}_{ij}(\nu=0)=\sum\limits_{m=1}^N (-1)^j\, {\bar q}_m^{\,i-1}
  \frac{{\bar p}_m}{\prod\limits_{s\neq m}^{N}
\,(q_m-q_{s})}\,\stackrel{m}{\sigma}_{j-1}(\bar\bfq)
 }
 \end{array}
 \eeq
In order to represent it as some linear combination of $S_{ab}$
(\ref{v304}) we use (\ref{v931}) or (\ref{v9313}). The choice
between these two possibilities comes from the requirement to have
the power of ${\bar q}$ in the interval $0...N$:
 \beq\label{v346}
 \begin{array}{c}
 \displaystyle{
L^{CG}_{ij}(\nu=0)=
 \left\{
 \begin{array}{l}
\sum\limits_{m=1}^N {\sum\limits_{c=0}^{N-j}} (-1)^{j+c}\, {\bar
q}_m^{\,i+c-1}
  \frac{{\bar p}_m}{\prod\limits_{s\neq m}^{N}
\,(q_m-q_{s})}\,{\sigma}_{j+c}(\bar\bfq)\,,\ \ j\geq i-1\,,
 \\
-\sum\limits_{m=1}^N \sum\limits_{c=0}^{j-1} (-1)^{j-c-1}\, {\bar
q}_m^{\,i-c-2}
  \frac{{\bar p}_m}{\prod\limits_{s\neq m}^{N}
\,(q_m-q_{s})}\,{\sigma}_{j-c-1}(\bar\bfq)\,,\ \ j< i-1\,,
 \end{array}
 \right.
 }
 \end{array}
 \eeq
In formula (\ref{v304}) the index of $\sigma(\bar\bfq)$ as well as
the power of ${\bar q}$ is an image of $\rho$-function. Therefore,
we should exclude somehow the terms corresponding to the indices
$N-1$ (they have no preimages). Dropping of terms with
$\sigma_{N-1}(\bar\bfq)$ does not change the sum since
$\sigma_{N-1}(\bar\bfq)=0$. This gives $(1-\delta_{c,N-j-1})$ in the
upper line of (\ref{v342}) and $(1-\delta_{j-c,N})$ in the lower
one. The exclusion of terms with ${\bar q}^{N-1}$ is not so simple.
These terms exist in the upper line of (\ref{v346}) for $j=i-1$ and
$j=i$. In the case $j=i-1$, ${\bar q}^{N-1}$ is set to zero since it
is multiplied by $\sigma_{N-1}(\bar\bfq)=0$. In the case $j=i$ all
such terms are equal to each other for all $i$ (with $c=N-i$ in the
sum), i.e. the terms with ${\bar q}^{N-1}$ form the scalar matrix
with the same diagonal elements value
$\sum\limits_{m=0}^N(-1)^N{\bar q}_m^{N-1}
  \frac{{\bar p}_m}{\prod\limits_{s\neq m}^{N}
\,(q_m-q_{s})}\,{\sigma}_{N}(\bar\bfq)$. Since $\tr {\bar P}=0$ then
$\tr L^{CG}(\nu=0)=0$. Thus, the sum of the unwanted term with the
rest of the sum should be traceless. This condition allows to
compute the unwanted term:
 \beq\label{v347}
 \begin{array}{c}
 \displaystyle{
\sum\limits_{m=1}^N(-1)^N{\bar q}_m^{N-1}
  \frac{{\bar p}_m{\sigma}_{N}(\bar\bfq)}{\prod\limits_{s\neq m}^{N}
\,(q_m-q_{s})}=\!-\frac{1}{N}
\sum\limits_{m=1}^N\sum\limits_{i=1}^N\sum\limits_{c=0}^{N-i-2}
(-1)^{i+c}\, {\bar q}_m^{\,i+c-1}
  \frac{{\bar p}_m{\sigma}_{i+c}(\bar\bfq)}{\prod\limits_{s\neq m}^{N}
\,(q_m-q_{s})}\,.
 }
 \end{array}
 \eeq
It provides the last one summand in (\ref{v342}):
 $$
-\delta_{i,j}\frac{1}{N}
\sum\limits_{k=1}^N\sum\limits_{c=0}^{N-k-2}S_{\varrho^{-1}(k+c-1),\,\varrho^{-1}(k+c)}=
-\delta_{i,j}\frac{1}{N}
\sum\limits_{k=1}^N\sum\limits_{c=0}^{N-k-2}S_{k+c,\,k+c+1}
 $$
 In the case $\nu\neq 0$ the computation is made in a similar way. In
 fact, the answer is the same as in $\nu=0$ case. The only remark -
 one should use ${\rm sl}_N$ bosonization formulae instead of ${\rm gl}_N$
 (\ref{v307}) since $\tr S=-N\nu$ (\ref{v20019}). The latter means that
  in the obtained formulae for $\nu=0$ we need to replace $S_{ab}\rightarrow
  S_{ab}-\delta_{ab}\frac{1}{N}\sum\limits_{l=1}^NS_{ll}$. It is
  easy to see that the Cartan part is contained in the line $j=i-1$.
  The corresponding correction is made in the upper line of
  (\ref{v342}). $\blacksquare$

}
     \renewcommand{\refname}{{\normalsize{References}}}


 \begin{small}

 \end{small}

\end{document}